\documentclass[onecolumn,aps,prd,showpacs,superscriptaddress,nofootinbib,amsmath,amssymb,floats,floatfix,showkeys,notitlepage,longbibliography]{revtex4-1}
\usepackage{comment}
\usepackage{xcolor}
\usepackage{graphicx}
\usepackage{caption}
\usepackage{subcaption}
\usepackage{palatino}
\usepackage[english]{babel}
\usepackage[commandnameprefix=always]{changes}
\usepackage{hyperref}
\hypersetup{colorlinks=true,linkcolor=blue,urlcolor=blue,citecolor=blue}
\usepackage[toc,page]{appendix}
\usepackage[normalem]{ulem}
\usepackage{orcidlink}
\usepackage{lipsum}
\usepackage{graphicx}
\usepackage{palatino}
\usepackage{sans}
\usepackage{adjustbox}
\usepackage{latexsym}
\usepackage{amsmath}
\usepackage{amssymb}
\usepackage{amsfonts}
\usepackage{dcolumn}
\usepackage{bm}
\usepackage{tikz}
\usepackage{bigints}
\usepackage{array,tabularx,multirow,booktabs}
\usepackage[tracking=true]{microtype}
\SetTracking{}{500}
\SetTracking{encoding={*}, shape=sc}{40}
\UseRawInputEncoding 
\allowdisplaybreaks
\usepackage{adjustbox}
\usepackage{latexsym}
\usepackage{amsmath}
\usepackage{amssymb}
\usepackage{amsfonts}
\usepackage{dcolumn}
\usepackage{bm}
\usepackage{tikz}
\usepackage{bigints}
\usepackage{array,tabularx,multirow,booktabs}
\usepackage[tracking=true]{microtype}
\usepackage{color}
\UseRawInputEncoding 
\allowdisplaybreaks

\begin{document} 

\title{Interplay of Null Energy Condition Violations and Thermodynamics in Kiselev Black Hole Evaporation}

\author{Vitalii Vertogradov}
\email{vdvertogradov@gmail.com}
\affiliation{Physics department, Herzen state Pedagogical University of Russia,
48 Moika Emb., Saint Petersburg 191186, Russia} 
\affiliation{Center for Theoretical Physics, Khazar University, 41 Mehseti Street, Baku, AZ-1096, Azerbaijan.}
\affiliation{SPB branch of SAO RAS, 65 Pulkovskoe Rd, Saint Petersburg
196140, Russia}

\author{Maksim Grigorev 
}
\email{maxgrig63@gmail.com}
\affiliation{Physics department, Herzen state Pedagogical University of Russia,
48 Moika Emb., Saint Petersburg 191186, Russia} 

\begin{abstract}
The evaporation of black holes with two horizons presents a rich thermodynamic landscape that departs fundamentally from the Schwarzschild paradigm. In this work, we analyze the Hawking temperature dynamics of the Kiselev black hole under varying mass $M$ and anisotropic fluid parameter $N$, explicitly connecting temperature behavior to phase transitions and violations of the null energy condition (NEC). We find that the temperature does not necessarily diverge during evaporation; instead, it typically falls to zero as the black hole evaporates. This cooling behavior is preceded, in certain parameter regimes, by a phase transition marked by a peak in temperature and a divergence in heat capacity. Crucially, the presence and nature of these phase transitions are dictated by the spacetime regions where the NEC is violated: global NEC violation leads to horizon merger and temperature suppression, while partial or absent violation can restore the standard evaporation picture. Our results establish a direct correspondence between thermodynamic stability, horizon dynamics, and energy condition structure in anisotropic black hole spacetimes.
\end{abstract}

\date{\today}

\keywords{Black hole; Thermodynamics; Kiselev spacetime; Energy Condition; Apparent Horizon.}

\pacs{95.30.Sf, 04.70.-s, 97.60.Lf, 04.50.Kd }

\maketitle
\section{Introduction}

Following the detection of gravitational waves from black hole mergers~\cite{bib:ligo1,bib:ligo2} and the observation of black hole shadows by the Event Horizon Telescope collaboration~\cite{bib:m87,bib:mw}, interest in black hole physics has surged dramatically.

One of the most remarkable discoveries in black hole theory is that black holes possess a temperature and can evaporate via Hawking radiation~\cite{bib:hawking}. In the standard picture of Schwarzschild black hole evaporation, the temperature diverges as the mass decreases, ultimately leading to the infamous information loss paradox. Another well-established result in black hole thermodynamics is that extremal black holes have zero temperature.

Most known exact black hole solutions-including Reissner--Nordstrom, Kerr, Kerr--Newman, and many regular black holes--feature two horizons in the non-extremal case. However, the evaporation dynamics of such two-horizon black holes differ fundamentally from the Schwarzschild case. As demonstrated in~\cite{bib:vertogradov2025plb}, a direct link exists between the null energy condition (NEC) and horizon behavior: when the NEC is violated, the outer horizon becomes a timelike hypersurface that shrinks, while the inner horizon becomes spacelike and expands. This evolution inevitably leads to the merger of the two horizons at a certain time, forming an extremal black hole with vanishing Hawking temperature.

To illustrate the temperature behavior during the evaporation of two-horizon black holes, we study the Kiselev metric~\cite{bib:kiselev}, which admits two horizons for non-exotic matter ($\omega > 0$) and reduces to the Reissner--Nordstrom solution for $\omega = 1/3$. We show that, during evaporation, the Hawking temperature may either (i) increase slightly to a maximum and then decrease to zero, or (ii) decrease monotonically to zero. Moreover, when one parameter is held fixed, the temperature evolution is directly governed by the heat capacity: it increases when the heat capacity is negative and decreases when it is positive. The Hawking temperature reaches its maximum precisely at a thermodynamic phase transition.

The Kiselev spacetime is analogous to the Husain solution~\cite{bib:husain}, both describing black holes surrounded by a anisotropic fluid obeying the barotropic equation of state $\bar{P} = \omega \rho$, where $\bar{P}$ denotes the averaged pressure. Although the thermodynamic properties of the Kiselev black hole have been extensively studied in the literature, all prior analyses have assumed static configurations with fixed parameters. In contrast, our work investigates accretion and evaporation processes in the quasi-static approximation to determine how the temperature evolves dynamically during black hole evaporation.

Dynamic generalizations of the Kiselev and Husain solutions have been widely employed to study gravitational collapse~\cite{bib:vertogradov2018ijmpa,bib:vertogradov2022ijmpa,bib:vertogradov2025cpc}. Notably, the Kiselev solution admits a homothetic Killing vector~\cite{bib:vertogradov2023mpla}, making it particularly suitable for analyzing the dynamical behavior of black hole shadows~\cite{bib:perlik,bib:vertogradov2024epjc,bib:ali2024plb}.

This paper is organized as follows. In Section~\ref{sec:kiselev_properties}, we review the general properties of the Kiselev metric. Section~\ref{sec:thermo} defines the key thermodynamic quantities of the Kiselev black hole and examines their dependence on the equation-of-state parameter $\omega$. Section~\ref{sec:temp_behavior} analyzes the temperature evolution during evaporation and accretion. In Section~\ref{sec:nec}, we establish the connection between temperature dynamics and the null energy condition. Finally, Section~\ref{sec:discussion} discusses our results and their implications.

Throughout this work, we adopt the metric signature $(-,+,+,+)$ and use geometrized units where $8\pi G = c = 1$.

\section{Kiselev black hole}
\label{sec:kiselev_properties}

The special type of the metric of the form
\begin{equation}
ds^2=-fdt^2+f^{-1}dr^2+r^2d\Omega^2
\end{equation}
is supported by an anisotropic fluid, which implies that the Einstein tensor satisfies the conditions $G^t_t = G^r_r$ and $G^\theta_\theta = G^\varphi_\varphi$. Correspondingly, the energy-momentum tensor exhibits the same symmetries: $T^t_t = T^r_r$ and $T^\theta_\theta = T^\varphi_\varphi$. These properties imply that the radial pressure $P_r$ must obey the equation of state $P_r = -\rho$. Consequently, assuming a tangential pressure governed by the equation of state
\begin{equation} \label{eq:eos1}
P_t = \alpha \rho,
\end{equation}
leads to the so-called Husain solution~\cite{bib:husain}, given by
\begin{equation}
f(r) = 1 - \frac{2M}{r} + \frac{N}{r^{2\alpha}}.
\end{equation}

However, the equation of state~\eqref{eq:eos1} is not barotropic in the conventional sense. This becomes evident when considering electromagnetic fields: instead of yielding the expected radiation-like equation of state at $\alpha = 1/3$, the correct electromagnetic solution arises at $\alpha = 1$. Similarly, the strong energy condition is violated not for $\alpha < -1/3$, but rather for $\alpha < 0$. These deviations are direct consequences of the intrinsic anisotropy of the energy-momentum tensor.

Attempting to impose isotropy on the energy-momentum tensor-i.e., requiring $P_r = P_t$ results in a highly restrictive scenario. The only consistent possibility under the condition $P = -\rho$ corresponds to the well-known Kottler solution, also known as the Schwarzschild--de Sitter spacetime.

To restore a more conventional barotropic framework while accounting for anisotropy, Kiselev~\cite{bib:kiselev} introduced the concept of an averaged pressure $\bar{P}$, defined as
\begin{equation}
\bar{P} = \frac{1}{3} \left( P_r + 2P_t \right).
\end{equation}
In this approach, the barotropic equation of state takes the form
\begin{equation} \label{eq:eos2}
P_t = \frac{1}{2}(3\omega + 1) \rho,
\end{equation}
and the solution to the Einstein equations yields the Kiselev metric:
\begin{equation} \label{eq:kiselev}
f(r) = 1 - \frac{2M}{r} + \frac{N}{r^{3\omega + 1}}.
\end{equation}

Here, the parameter $N$ can be interpreted as a combination of electric and magnetic charges of a black hole supported by nonlinear electrodynamics, particularly when $\omega > 0$~\cite{bib:vertogradov2025non}. Notably, in this formulation, the value $\omega = 1/3$ correctly reproduces the Reissner--Nordstrom solution, in contrast to the earlier parametrization. 

An important feature of the Kiselev solution~\eqref{eq:kiselev} is that the weak energy conditions require $N\omega > 0$, implying that the parameters $N$ and $\omega$ must have the same sign. This condition ensures the physical consistency of the matter source and highlights the interplay between geometry and the underlying anisotropic fluid model.

Our goal is to investigate the thermodynamic properties of Kiselev's solution \eqref{eq:kiselev} as the parameters $M$ and $N$ vary. To achieve this, we first need to determine the positions of the horizons. However, for the general case with $\omega \in [0,1]$, it is not possible to explicitly find the roots of the equation. Instead, we will first show that the equation $f(r) = 0$ has at most two positive real roots. This will demonstrate that the black hole described by \eqref{eq:kiselev} possesses at most two horizons.

To show that the equation
\[
r^{3\omega+1} - 2M r^{3\omega} + N = 0
\]
has at most two positive real roots under the conditions \( M > 0 \), \( N > 0 \), and \( 0 < \omega <1 \) let us perform the change of variable: let \( x = r^{3\omega} \) and  define \( \alpha = 1 + \frac{1}{3\omega} \). Since \( 0 < \omega < 1 \), we have \( 3\omega \in (0,3) \), so \( \frac{1}{3\omega} > \frac{1}{3} \), and thus \( \alpha > 1 + \frac{1}{3} = \frac{4}{3} > 1 \).

The equation then becomes:
\[
x^{\alpha} - 2M x + N = 0, \quad x > 0.
\]
Define the function
\[
g(x) = x^{\alpha} - 2M x + N, \quad x > 0.
\]
Our goal is to show that the equation \( g(x) = 0 \) has at most two positive real solutions.

Compute the first and second derivatives:
\[
g'(x) = \alpha x^{\alpha - 1} - 2M,
\]
\[
g''(x) = \alpha(\alpha - 1) x^{\alpha - 2}.
\]
Since \( \alpha > 1 \) and \( x > 0 \), we have \( \alpha(\alpha - 1) > 0 \) and \( x^{\alpha - 2} > 0 \), so \( g''(x) > 0 \) for all \( x > 0 \). This means that \( g(x) \) is \textbf{strictly convex} (convex downward) on \( (0, \infty) \).

Now recall the following classical result from calculus:

\medskip

\textbf{Rolle's Theorem.} \textit{Let \( h \) be a function that is continuous on the closed interval \( [a, b] \), differentiable on the open interval \( (a, b) \), and satisfies \( h(a) = h(b) = 0 \). Then there exists a point \( c \in (a, b) \) such that \( h'(c) = 0 \).}

\medskip

Apply this theorem to \( g(x) \). Suppose, for contradiction, that \( g(x) = 0 \) has three distinct positive roots: \( 0 < x_1 < x_2 < x_3 \). Then \( g(x_1) = g(x_2) = g(x_3) = 0 \).

By Rolle's Theorem applied on \( [x_1, x_2] \), there exists \( \xi_1 \in (x_1, x_2) \) such that \( g'(\xi_1) = 0 \).

Similarly, on \( [x_2, x_3] \), there exists \( \xi_2 \in (x_2, x_3) \) such that \( g'(\xi_2) = 0 \).

Now, \( g'(x) \) is differentiable on \( (0, \infty) \), and \( g'(\xi_1) = g'(\xi_2) = 0 \). Applying Rolle's Theorem to \( g'(x) \) on \( [\xi_1, \xi_2] \), we conclude that there exists \( \eta \in (\xi_1, \xi_2) \) such that \( g''(\eta) = 0 \).

But this contradicts the fact that \( g''(x) > 0 \) for all \( x > 0 \). Hence, the assumption of three positive roots is false.

Therefore, the equation \( g(x) = 0 \) has at most two positive real solutions. Since the mapping \( r \mapsto x = r^{3\omega} \) is strictly increasing and bijective from \( (0,\infty) \) to \( (0,\infty) \), the number of positive real roots in \( r \) is the same as in \( x \).

Thus, we have shown that the Kiselev black hole for $\omega > 0$ possesses at most two horizons. Next, we determine the relationship between the parameters $M$ and $N$ under which the Kiselev black hole has two distinct horizons. Using the extremality condition $f(r_h) = f'(r_h) = 0$, we find the horizon radius
\begin{equation}
r_h = \frac{6\omega M}{3\omega + 1}.
\end{equation}
Substituting this value of the extremal horizon radius into the Kiselev metric \eqref{eq:kiselev}, we obtain the following inequality:
\begin{eqnarray} \label{eq:ineq}
N \leq \left(2 - \frac{2\zeta}{\zeta + 1}\right) \left(\frac{2\zeta}{\zeta + 1}\right)^\zeta M^{\zeta + 1}, \nonumber \\
\zeta \equiv 3\omega, \quad 2 - \frac{2\zeta}{\zeta + 1} > 0.
\end{eqnarray}
If the inequality is strict ($<$), the Kiselev black hole has two distinct horizons. If equality holds, the black hole is extremal. If the inequality is violated, no horizons exist and the spacetime contains a naked singularity.

As a notable example, consider the case $\omega = \frac{1}{3}$, in which the Kiselev metric \eqref{eq:kiselev} reduces to the Reissner--Nordstrom metric, with $N \equiv Q^2$, where $Q$ is the electric charge of the black hole. In this case, $\zeta = 1$, and we recover the well-known condition for the Reissner--Nordstrom solution: $Q^2 \leq M^2$.

\section{Thermodynamic Properties of the Kiselev Solution}
\label{sec:thermo} 
As established in the previous section, for the equation-of-state parameter $\omega \in (0,1]$, the Kiselev metric~\eqref{eq:kiselev} possesses at most two horizons: one in the extremal case, and none when inequality~\eqref{eq:ineq} is violated. For arbitrary values of $\omega$, the horizon locations cannot be determined analytically. Consequently, we resort to approximate and numerical methods.

The lapse function of the Kiselev solution can be expressed as
\begin{equation}
    f(r) = 1 - \frac{b(r)}{r},
\end{equation}
where $b(r)$ is interpreted as the shape function~\cite{bib:dirty}. For the Kiselev metric, this function takes the form
\begin{equation}
    b(r) = 2M - N r^{1 - 2\alpha},
\end{equation}
with $\alpha = \frac{3\omega + 1}{2}$, so that $1 - 2\alpha = -3\omega$.

The Hawking temperature is evaluated at the outer event horizon $r_+$ and is proportional to the surface gravity $\kappa$:
\begin{equation}
    T = \frac{\kappa}{2\pi}.
\end{equation}
The surface gravity, in terms of the shape function, is given by
\begin{equation}
    \kappa = \frac{1}{2r_{+}}\left(1 - b'(r_+)\right).
\end{equation}
Substituting $b'(r) = -3\omega N r^{-3\omega - 1}$ yields the Hawking temperature for the Kiselev black hole:
\begin{equation} \label{eq:temperature}
    T_{H} = \frac{1}{4\pi r_{+}} \left(1 - \frac{3\omega N}{r_+^{3\omega+1}}\right).
\end{equation}

In addition to temperature, we examine other thermodynamic quantities, such as the Bekenstein–Hawking entropy,
\begin{equation}
    S = \pi r_{+}^2,
\end{equation}
and the heat capacity at constant parameters,
\begin{equation}
    C = T_{H}\frac{dS}{dT_{H}} = T_{H}\frac{dS}{dr_{+}}\left(\frac{dT_{H}}{dr_{+}}\right)^{-1}.
\end{equation}
For the Kiselev metric, the heat capacity becomes
\begin{equation} \label{eq:phase}
    C = -2\pi r_{+}^{2}\,\frac{1 - 3\omega N r_+^{-3\omega-1}}{1 - (9\omega^2 + 6\omega)N r_+^{-3\omega-1}}.
\end{equation}

A divergence in the heat capacity signals a thermodynamic phase transition. From expression~\eqref{eq:phase}, such a transition occurs when the denominator vanishes, i.e., when
\begin{equation} \label{eq:phase1}
    N_{\mathrm{PT}} = \frac{r_{+}^{3\omega+1}}{9\omega^2 + 6\omega}.
\end{equation}
This critical value can also be rewritten in terms of the black hole mass $M$. Using the horizon condition $f(r_+) = 0$, namely $2M = r_+ + N r_+^{-3\omega}$, one obtains
\begin{equation}
    N_{\mathrm{PT}} = \left( \frac{2M\,(9\omega^2 + 6\omega)^{\frac{3\omega}{3\omega+1}}}{(3\omega + 1)^2} \right)^{3\omega + 1}.
\end{equation}

In the special case $\omega = \tfrac{1}{3}$ (corresponding to radiation or a Maxwell field, with $N = Q^2$), these expressions reduce to
\[
    Q_{\mathrm{PT}} = \frac{r_+}{\sqrt{3}}, \qquad Q_{\mathrm{PT}} = \frac{\sqrt{3}}{2}\,M,
\]
which precisely match the well-known Reissner–Nordstrom results for the charge at which a phase transition occurs.

We now proceed to analyze the behavior of the outer horizon radius $r_+(\omega)$ using numerical methods. These techniques yield approximate values of $r_+$ over the interval $\omega \in (0,1]$.

With knowledge of $r_+(\omega)$-the maximal horizon radius for a given $\omega$-we can further investigate the heat capacity as a function of $\omega$.
\begin{figure}[ht]
    \centering
    \includegraphics[width=0.6\linewidth]{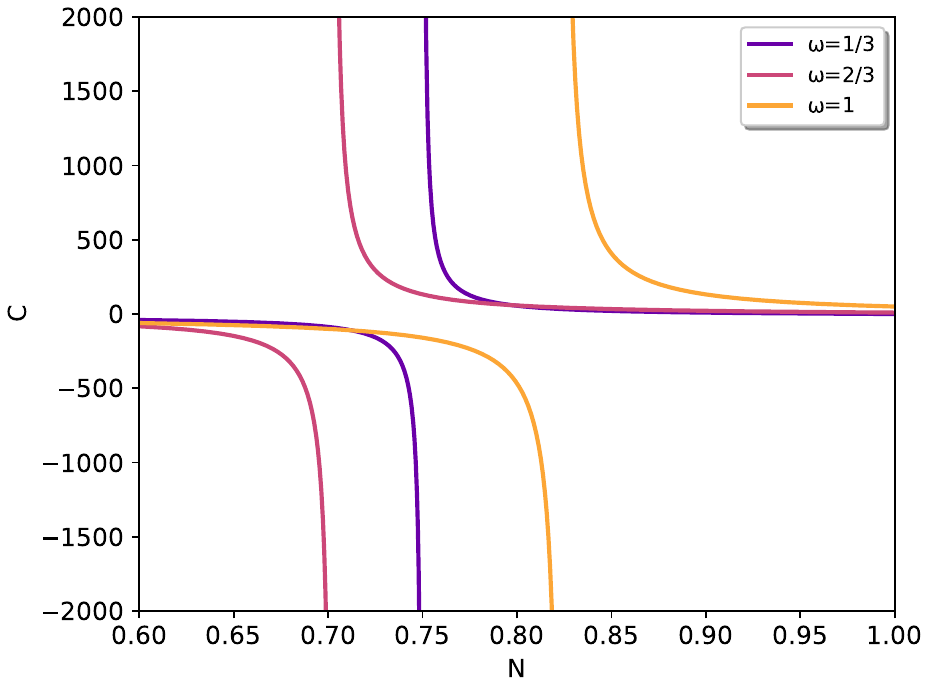}
    \caption{With $N = 0.6,\, 0.65,\,0.75,\, 0.85,\, 0.95$, phase transitions are observed for alpha values between 0.7 and 1, marking a shift from black hole instability to local stability. A further phase transition occurs as alpha approaches 2, corresponding to a switch back from local stability to instability.}
    \label{fig:enter-label}
\end{figure}
Now we aim to determine the function $T(r_{\mathrm{pt}})$, where $r_{\mathrm{pt}}$ denotes the event horizon radius at which a thermodynamic phase transition occurs. To this end, we evaluate the lapse function under phase-transition conditions. Substituting the critical value $N = N_{\mathrm{PT}}$ from Eq.~\eqref{eq:phase1} into the Kiselev metric yields
\begin{equation}
    f(r_{\mathrm{pt}}) = 1 - \frac{2M}{r_{\mathrm{pt}}} + \frac{1}{9\omega^2 + 6\omega} = 0.
\end{equation}
Solving this equation for the phase-transition radius gives
\begin{equation}
    r_{\mathrm{pt}} = \frac{2M(9\omega^2 + 6\omega)}{(3\omega + 1)^2}.
\end{equation}
In the specific case $\omega = \tfrac{1}{3}$, this expression reduces to the well-known Reissner--Nordstrom phase-transition radius expressed in terms of the black hole mass.

\begin{figure}[ht]
    \centering
    \begin{subfigure}[t]{0.47\textwidth}
        \includegraphics[width=\linewidth]{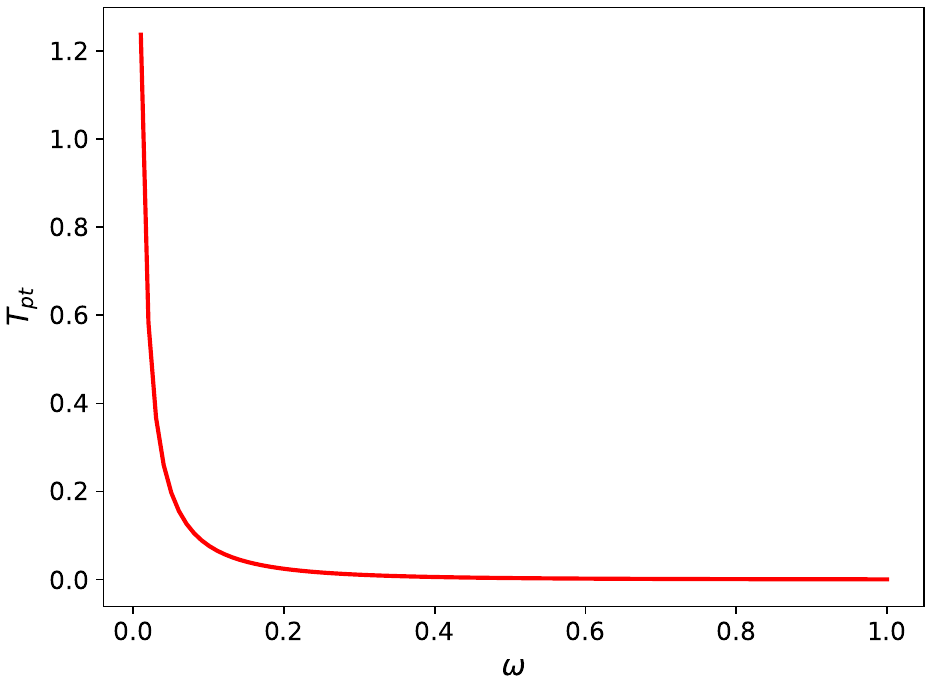}
        \caption{}
        \label{fig:nc}
    \end{subfigure}
    \hfill
    \begin{subfigure}[t]{0.47\textwidth}
        \includegraphics[width=\linewidth]{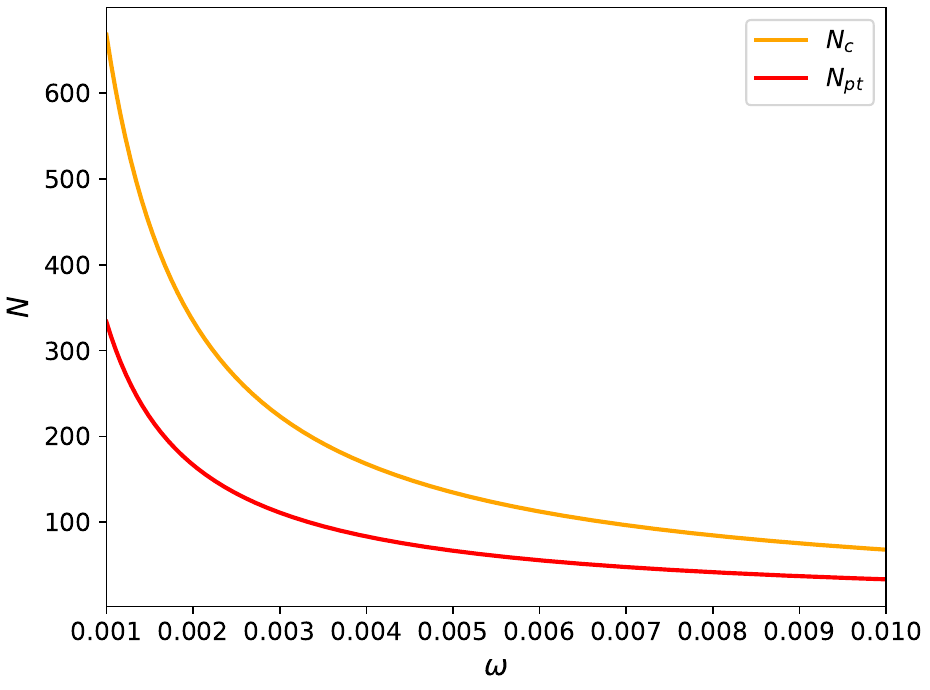}
        \caption{}
        \label{fig:npt}
    \end{subfigure}
    \caption{%
        (a) Temperature of phase transitions as a function of $\omega$. The temperature increases as $\omega$ decreases. 
        (b) Comparison between $N_{\mathrm{PT}}$ and $N_{c}$. For $\omega= 0$, the ratio $N_{c}/N_{\mathrm{PT}}$ is approximately 2.
    }
    \label{fig:graphs}
\end{figure}

Another relation of interest is the inequality linking the black hole mass $M$ and the parameter $N$:
\begin{equation}
    M \geq \left(3\omega N\right)^{\frac{-3\omega}{3\omega+1}} \left(\frac{3\omega+1}{2}\right) N.
\end{equation}
Equality holds precisely for an extremal black hole. To assess how close the phase-transition parameter $N_{\mathrm{PT}}$ is to the extremal value $N_c$, we first compute the horizon radius of the extremal configuration. This is obtained by solving $f(r) = 0$ and $f'(r) = 0$ simultaneously, yielding
\begin{equation}
    r_{\mathrm{ext}} = \left( \frac{(3\omega + 1) N}{2M} \right)^{\frac{1}{3\omega}}.
\end{equation}

We now eliminate $M$ between the extremality condition and the phase-transition condition. Solving the extremality relation for $M$ and substituting into the definition of $N_{\mathrm{PT}}$ leads to the critical parameter for the extremal (or ``critical'') black hole:
\begin{equation}
    N_{c} = \frac{r_{+}^{3\omega + 1}}{3\omega}.
\end{equation}
In the limit $\omega \to 0$, the phase-transition parameter approaches half the extremal value, i.e., $N_{\mathrm{PT}} \to \tfrac{1}{2} N_{c}$, as illustrated in Fig.~\ref{fig:npt}.
\section{Temperature Behavior}
\label{sec:temp_behavior} 
During the evaporation of a Schwarzschild black hole, the Hawking temperature increases as the mass decreases. However, the presence of two event horizons significantly complicates the evaporation dynamics. For the Kiselev black hole with $\omega \in (0, 1]$, two horizons generally exist, and the temperature depends on both the mass $M$ and the parameter $N$. As these parameters evolve-particularly during evaporation-the temperature exhibits nontrivial behavior. Moreover, the evaporation process violates the null energy condition, which, as shown in~\cite{bib:vertogradov2025plb}, causes the outer horizon to become a timelike hypersurface that shrinks, while the inner horizon becomes a spacelike hypersurface that expands. Eventually, the two horizons merge at specific values of $M$ and $N$, forming an extremal black hole with vanishing temperature. To systematically study this behavior, we consider three distinct scenarios:
\begin{enumerate}
    \item $M$ varies while $N$ remains constant,
    \item $N$ varies while $M$ remains constant,
    \item Both $M$ and $N$ vary simultaneously.
\end{enumerate}

\subsection{Mass Variation at Constant $N$}
\label{subsec:M_variation}

We assume the black hole mass evolves from an initial value $M_0$, either decreasing due to Hawking evaporation or increasing via accretion. Consider an infinitesimal change in mass:
\begin{equation}
M = M_0 + \beta M_0, \quad |\beta| \ll 1,
\end{equation}
where $\beta > 0$ corresponds to accretion and $\beta < 0$ to evaporation. The event horizon radius $r_+$ is fully determined by $M$ and $N$. During evaporation ($\beta < 0$), the horizon shrinks; during accretion ($\beta > 0$), it expands. Thus, the derivative of the horizon radius with respect to mass satisfies
\begin{eqnarray}
\frac{\partial r_+}{\partial M} &>& 0, \quad \text{(accretion)}, \nonumber \\
\frac{\partial r_+}{\partial M} &<& 0, \quad \text{(evaporation)}.
\end{eqnarray}

The Hawking temperature under this variation reads
\begin{equation} \label{eq:temp_case1}
T_H\left(M_0+\beta M_0, N\right) = \frac{1}{4\pi r_+\left(M_0+\beta M_0, N\right)} \left[1 -\frac{3\omega N}{r_+^{3\omega+1}\left(M_0+\beta M_0, N\right)} \right].
\end{equation}

To analyze the temperature response, we expand $T_H$ in a Taylor series in $\beta$:
\begin{equation}
T_H = T_H^{(0)} + \beta T_H^{(1)} + \mathcal{O}(\beta^2),
\end{equation}
where $T_H^{(0)}$ is the initial temperature (cf. Eq.~\eqref{eq:temperature}) and the first-order correction is
\begin{equation}
T_H^{(1)} = \frac{M_0}{4\pi\left(r_+^{(0)}\right)^{3\omega+3}} \frac{\partial r_+}{\partial M}\bigg|_{M = M_0} \left[\left(9\omega^2 - 6\omega\right) N - \left(r_+^{(0)}\right)^{3\omega+1}\right].
\end{equation}
Here, $r_+^{(0)} \equiv r_+(M_0, N)$ denotes the initial horizon radius.

The sign of $T_H^{(1)}$ is governed by the bracketed term. Its zero corresponds precisely to the phase-transition condition~\eqref{eq:phase1}. Consequently, we observe the following qualitative behaviors:
\begin{itemize}
\item During accretion, the temperature either monotonically decreases (negative heat capacity) or initially increases slightly (positive heat capacity near extremality) before decreasing.
\item During evaporation, the temperature either monotonically decreases to zero (positive heat capacity near extremality) or first increases (negative heat capacity), reaches a maximum at the phase transition, and then decreases to zero (positive heat capacity).
\end{itemize}
\begin{figure}[ht]
    \centering
    \begin{subfigure}[ht]{0.49\textwidth}
        \includegraphics[width=\linewidth]{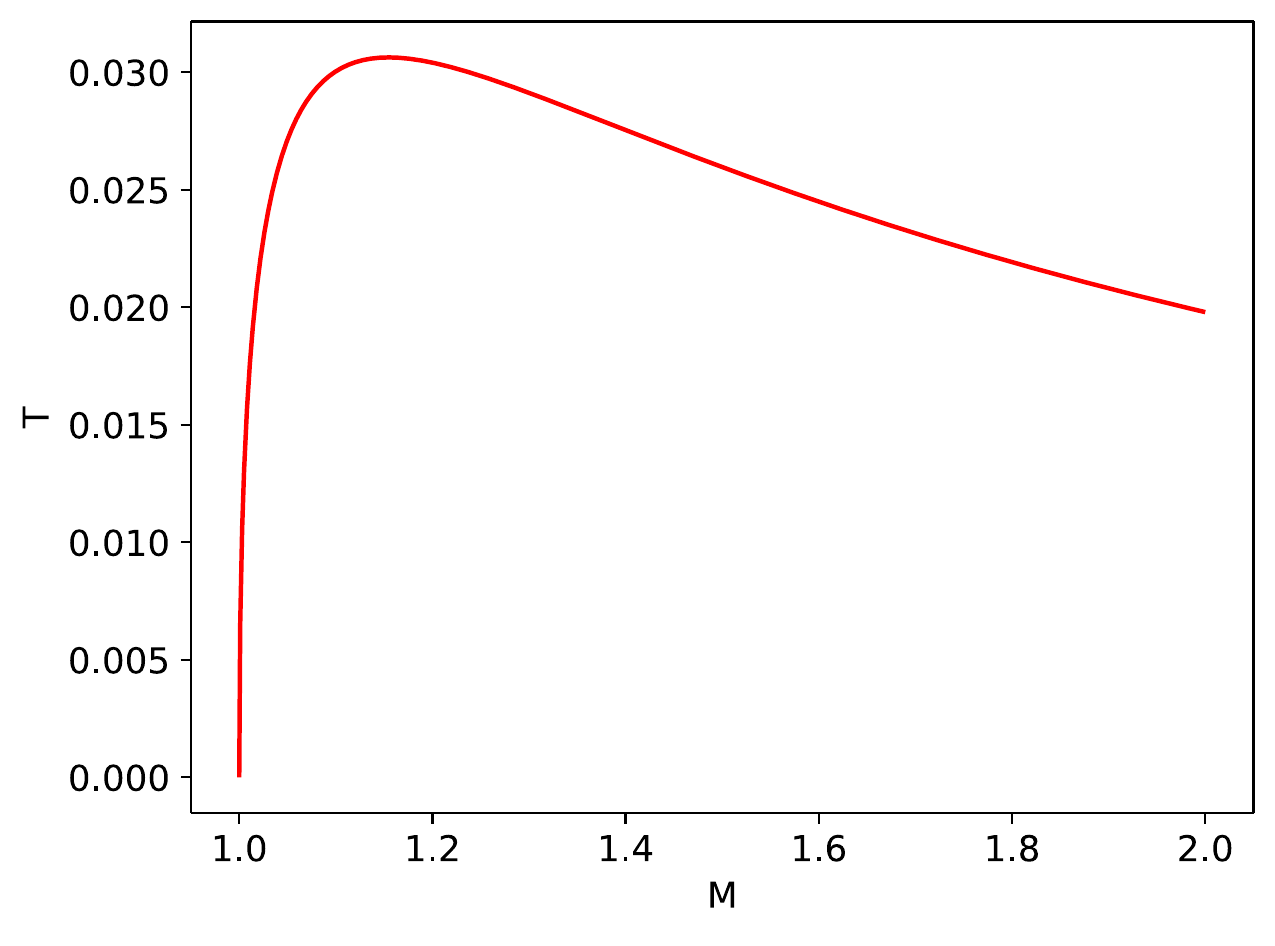}
        \caption{}
        \label{fig:incM}
    \end{subfigure}
    \hfill
    \begin{subfigure}[ht]{0.49\textwidth}
        \includegraphics[width=\linewidth]{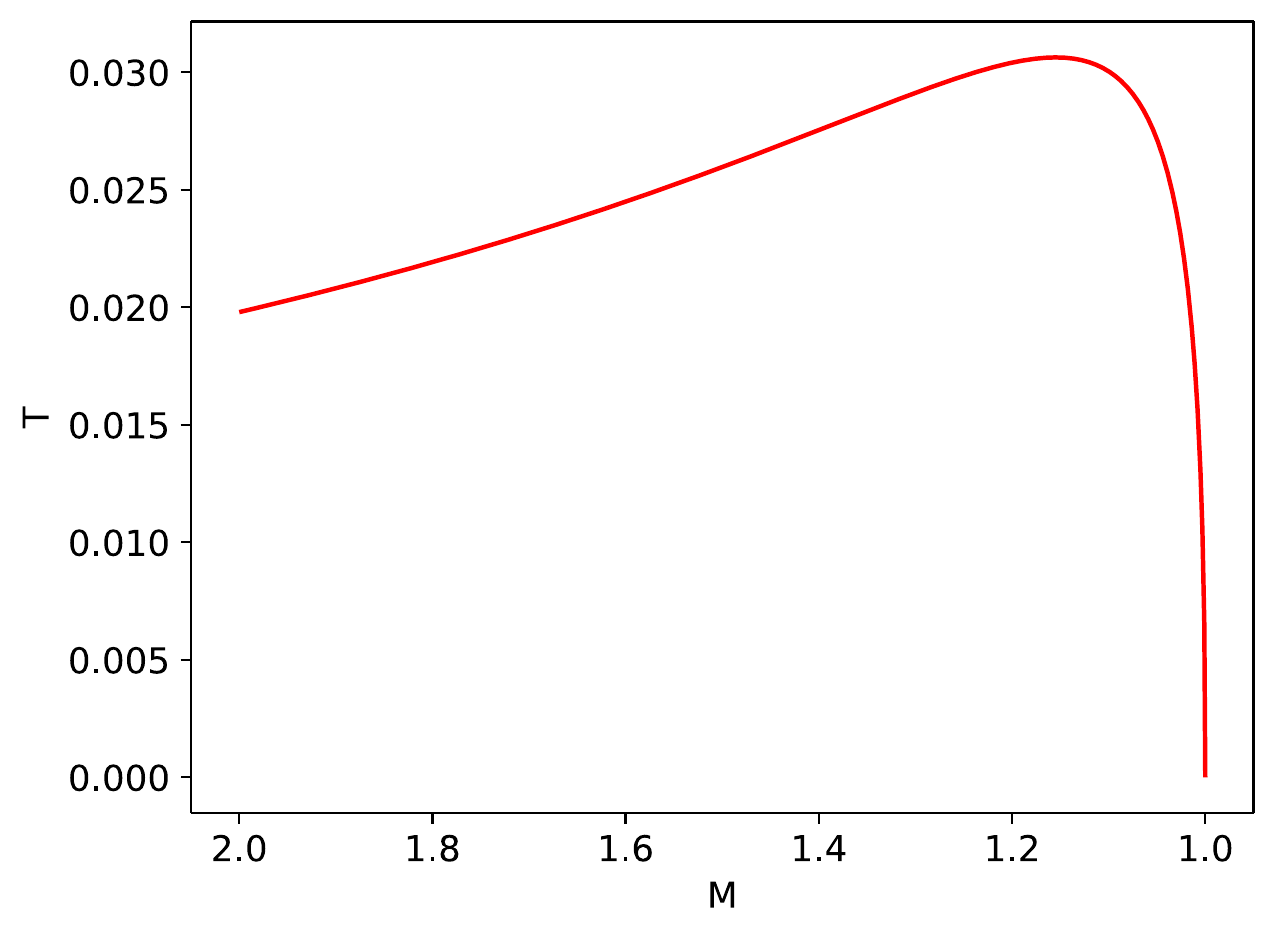}
        \caption{}
        \label{fig:decM}
    \end{subfigure}
    \caption{%
        (a) Temperature as a function of mass for increasing $M$ at fixed $N = 1$. The temperature starts from zero (extremal limit, $M = N$), rises to a maximum (phase transition), and then decreases monotonically.
        (b) Temperature for decreasing $M$-the mirror image of (a)-showing a monotonic rise to a critical maximum followed by a sharp drop to zero.
    }
    \label{fig:mass_variation}
\end{figure}

\subsection{$N$ Variation at Constant $M$}
\label{subsec:N_variation}

We now examine the case where $N$ varies while the mass $M$ is held fixed. Introducing a small perturbation,
\begin{equation}\label{eq:temp_case2}
    N = N_0 + \gamma N_0, \quad |\gamma| \ll 1,
\end{equation}
with $\gamma > 0$ representing accretion of the exotic field and $\gamma < 0$ its evaporation. The horizon radius is now a function of $N_0$ and $M$, and we assume
\begin{eqnarray}
\frac{\partial r_+}{\partial N} &>& 0, \quad \text{(accretion)}, \nonumber \\
\frac{\partial r_+}{\partial N} &<& 0, \quad \text{(evaporation)}.
\end{eqnarray}

The temperature becomes
\begin{equation}
T_H\left(M, N_0+\gamma N_0\right) = \frac{1}{4\pi r_+\left(M, N_0+\gamma N_0\right)} \left[1 -\frac{3\omega (N_0 + \gamma N_0)}{r_+^{3\omega+1}\left(M, N_0+\gamma N_0\right)} \right].
\end{equation}

Expanding to first order in $\gamma$,
\begin{equation}
    T_H = T_H^{(0)} + \gamma T_H^{(1)} + \mathcal{O}(\gamma^2),
\end{equation}
with the linear correction given by
\begin{equation}
    T_H^{(1)} = -\frac{N_0}{4\pi \left(r_+^{(0)}\right)^{3\omega+3}}\left[3\omega r^{(0)}_+ - \frac{\partial r_+}{\partial N}\bigg|_{N=N_0}\left((9\omega^2+6\omega)N_0 - \left(r_+^{(0)}\right)^{3\omega+1}\right)\right].
\end{equation}

Again, the sign of $T_H^{(1)}$ is determined by the expression in brackets. Setting it to zero yields the critical condition
\begin{equation}
\frac{\partial r_+}{\partial N}\bigg|_{N=N_0} = \frac{3\omega r^{(0)}_+}{(9\omega^2+6\omega)N_0 - \left(r_+^{(0)}\right)^{3\omega+1}}.
\end{equation}

Notably, for $\omega = 1/3$ (Reissner-Nordstrom limit), this condition is satisfied only when $N_0 = 0$, i.e., when the black hole reduces to the Schwarzschild case. Thus:

\begin{itemize}
\item During accretion ($\gamma > 0$), the temperature decreases monotonically from its Schwarzschild maximum toward zero (extremal limit), reflecting negative heat capacity throughout.
\item During evaporation ($\gamma < 0$), the temperature increases toward the Schwarzschild value. The only point where the derivative vanishes is at $N = 0$, corresponding to a phase transition from a Reissner–Nordstrom-like configuration to the Schwarzschild black hole.    
\end{itemize}

\begin{figure}[ht]
    \centering
    \begin{subfigure}[ht]{0.49\textwidth}
        \includegraphics[width=\linewidth]{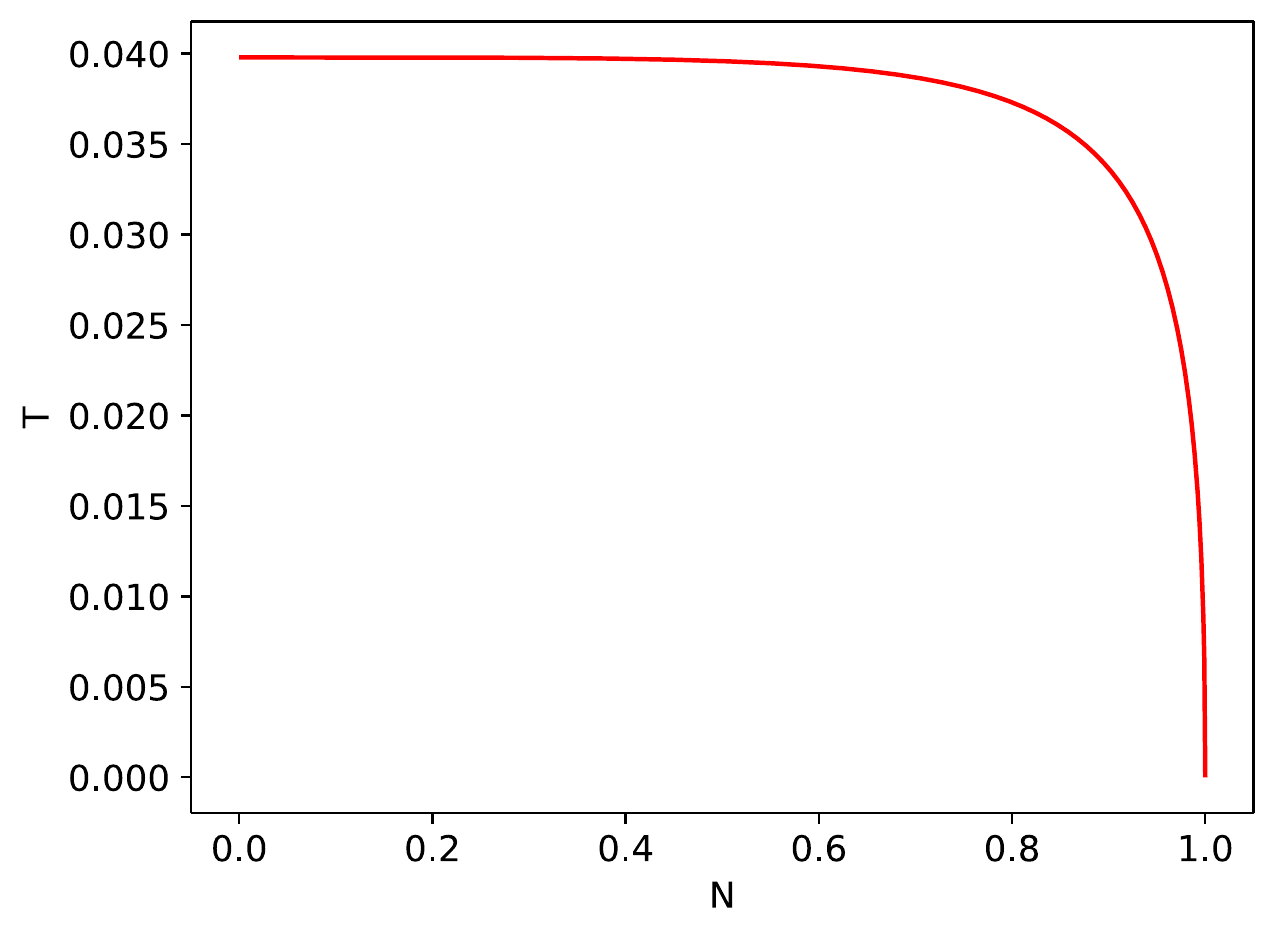}
        \caption{}
        \label{fig:incN}
    \end{subfigure}
    \hfill
    \begin{subfigure}[ht]{0.49\textwidth}
        \includegraphics[width=\linewidth]{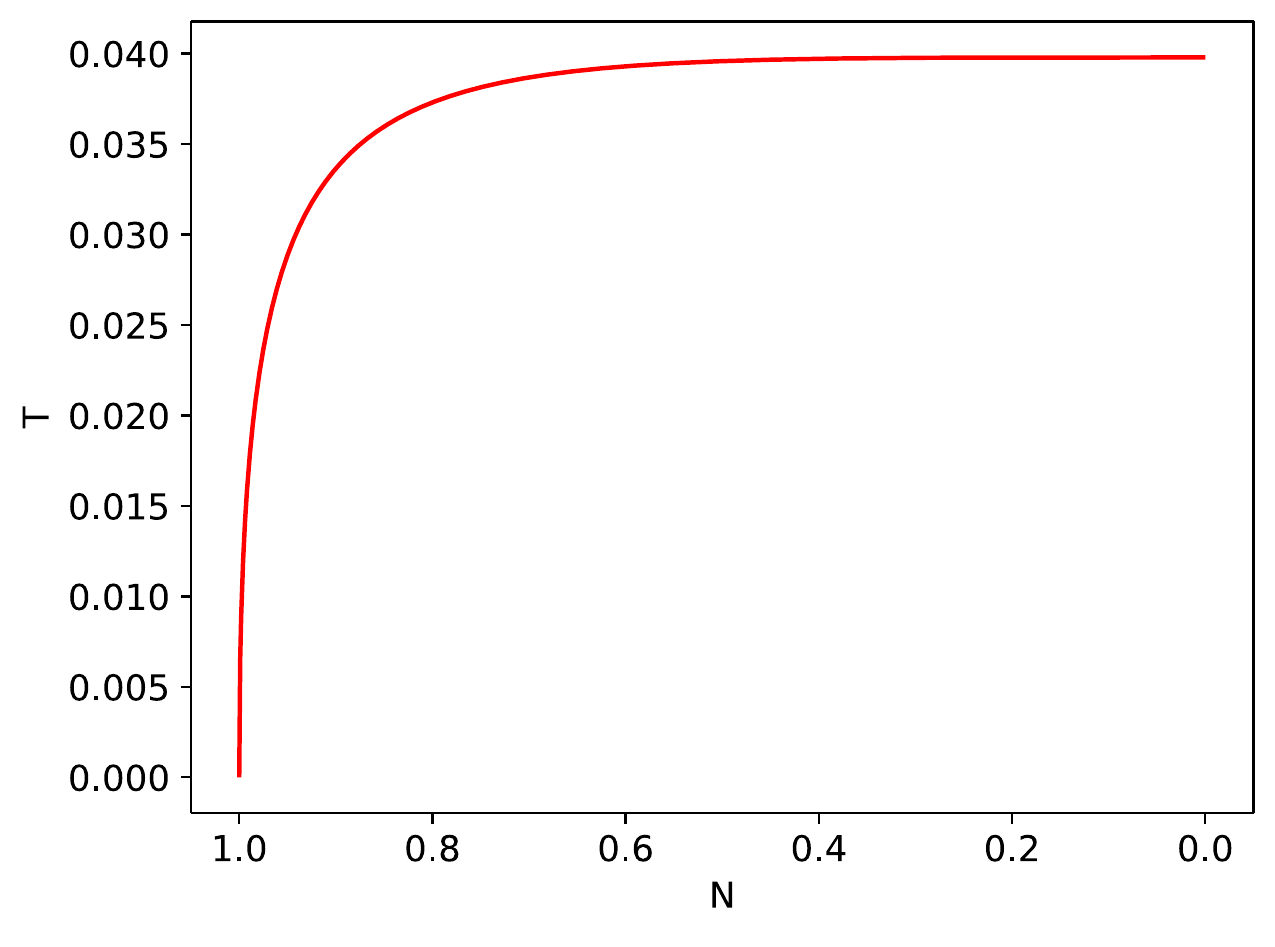}
        \caption{}
        \label{fig:decN}
    \end{subfigure}
    \caption{%
        (a) Temperature as $N$ increases at fixed $M = 1$. The curve starts at the Schwarzschild maximum ($N=0$), where the derivative vanishes-indicating a phase transition-and then decreases parabolically to zero (extremal limit, $M = N$).
        (b) Temperature as $N$ decreases-mirror image of (a)-showing a rise toward the Schwarzschild limit.
    }
    \label{fig:N_variation}
\end{figure}
\subsection{Simultaneous Variation of $N$ and $M$}

We now consider the general scenario in which both parameters $M$ and $N$ vary simultaneously. In this case, the Hawking temperature becomes a function of the two dimensionless perturbation parameters $\beta$ and $\gamma$, introduced in Eqs.~\eqref{eq:temp_case1} and~\eqref{eq:temp_case2}, respectively. The temperature is fully determined by $\beta$, $\gamma$, and the initial values $M_0$, $N_0$:

\begin{equation}
\begin{split}
    T_H\left(M_0+\beta M_0,\, N_0+\gamma N_0\right) = \frac{1}{4\pi r_+\left(M_0+\beta M_0,\, N_0+\gamma N_0\right)} \left[1 -\frac{3\omega (N_0 + \gamma N_0)}{r_+^{3\omega+1}\left(M_0+\beta M_0,\, N_0+\gamma N_0\right)} \right].
\end{split}
\end{equation}

Expanding to first order in the small parameters $\beta$ and $\gamma$, we write
\begin{equation}
    T_H = T_H^{(0)} + T_H^{(1)}(\beta, \gamma) + \mathcal{O}(\beta^2, \gamma^2, \beta\gamma),
\end{equation}
where the linear correction is given by
\begin{equation}
\begin{split}
T_H^{(1)}(\beta, \gamma) = \frac{1}{4\pi \left(r^{(0)}_+\right)^{3 + 3\omega}} 
\Bigg[ & \beta M_0 \frac{\partial r_+}{\partial M} 
\left(
(9\omega^2 + 6\omega)N_0 - \left(r^{(0)}_+\right)^{3 \omega+1}
\right) \\
& - \gamma N_0 
\left(
3\omega r^{(0)}_+ - \frac{\partial r_+}{\partial N} 
\left[
(9\omega^2 + 6\omega)N_0 - \left(r^{(0)}_+\right)^{3 \omega+1}
\right]
\right)
\Bigg].
\end{split}
\end{equation}

As in the previous cases, the sign of $T_H^{(1)}$-and thus whether the temperature initially increases or decreases-is governed by the expression in the large square brackets.

This two-parameter scenario naturally generalizes the earlier limiting cases: setting $\gamma = 0$ recovers the mass-variation analysis (Section~\ref{subsec:M_variation}), while $\beta = 0$ reproduces the $N$-variation results (Section~\ref{subsec:N_variation}).

To locate the onset of a phase transition in this generalized setting, we set the bracketed expression to zero. Introducing $\alpha = \frac{3\omega + 1}{2}$ for notational compactness, so that $3\omega = 2\alpha - 1$ and $9\omega^2 + 6\omega = 4\alpha^2 - 1$, we define
\begin{equation}
    D \equiv (4\alpha^2 - 1)N_0 - \left(r^{(0)}_+\right)^{2\alpha}.
\end{equation}
The phase-transition condition then becomes
\begin{equation}
\beta M_0 \frac{\partial r_+}{\partial M} D + \gamma N_0 \frac{\partial r_+}{\partial N} D = \gamma N_0 (2\alpha - 1) r^{(0)}_+,
\end{equation}
which can be solved for $\beta$ as a function of $\gamma$:
\begin{equation}
\beta = \gamma \cdot \frac{N_0 \left[(2\alpha - 1) r^{(0)}_+ - \frac{\partial r_+}{\partial N} D \right]}{M_0 \frac{\partial r_+}{\partial M} D}.
\label{eq:beta_from_gamma}
\end{equation}

This relation reveals a key physical insight: phase transitions can occur not only when the mass grows faster than the charge-like parameter $N$, but also in the opposite regime-when $N$ increases more rapidly than $M$. This is particularly striking because, as shown in Section~\ref{subsec:N_variation}, varying $N$ alone (at fixed $M$) does not produce a phase transition for generic $\omega$; the interplay between simultaneous changes in both parameters is essential.

Denoting the proportionality factor in Eq.~\eqref{eq:beta_from_gamma} by $K$, i.e.,
\begin{equation}
    \beta = \gamma \cdot K, \quad K = \frac{N_0 \left[(2\alpha - 1) r^{(0)}_+ - \frac{\partial r_+}{\partial N} D \right]}{M_0 \frac{\partial r_+}{\partial M} D},
\end{equation}
we can characterize the relative rates of change:

\begin{align}
&\text{Mass increases faster than } N \ (\beta > \gamma): \quad 
\begin{cases}
K > 1 & \text{if } \gamma > 0, \\
K < 1 & \text{if } \gamma < 0,
\end{cases} \\
&N \text{ increases faster than mass } (\gamma > \beta): \quad 
\begin{cases}
K < 1 & \text{if } \gamma > 0, \\
K > 1 & \text{if } \gamma < 0.
\end{cases}
\end{align}

These conditions delineate distinct thermodynamic regimes, each with qualitatively different temperature evolution.

\begin{figure}[ht]
    \centering
    \begin{subfigure}[t]{0.47\textwidth}
        \includegraphics[width=\linewidth]{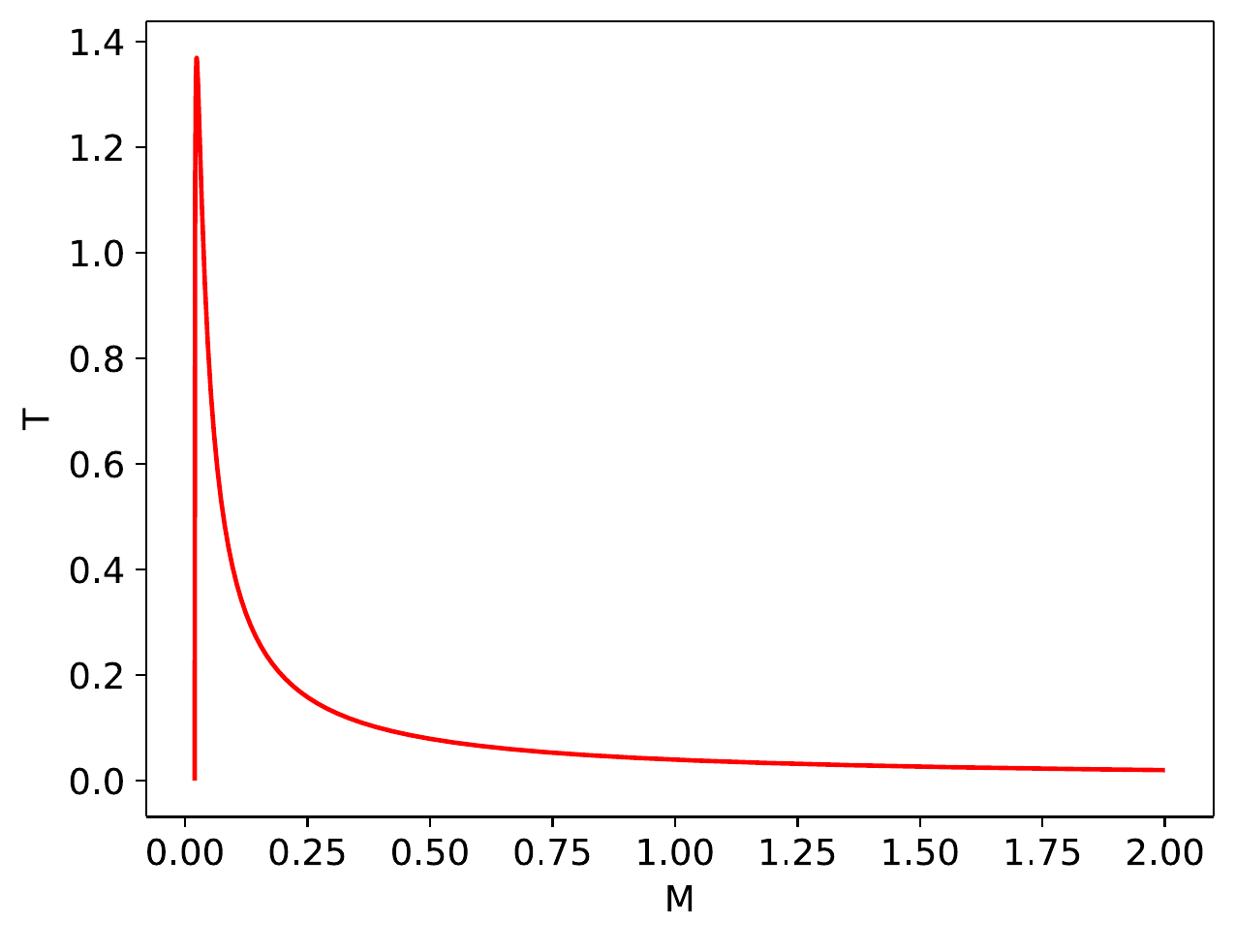}
        \caption{}
        \label{fig:M_faster}
    \end{subfigure}
    \hfill
    \begin{subfigure}[t]{0.49\textwidth}
        \includegraphics[width=\linewidth]{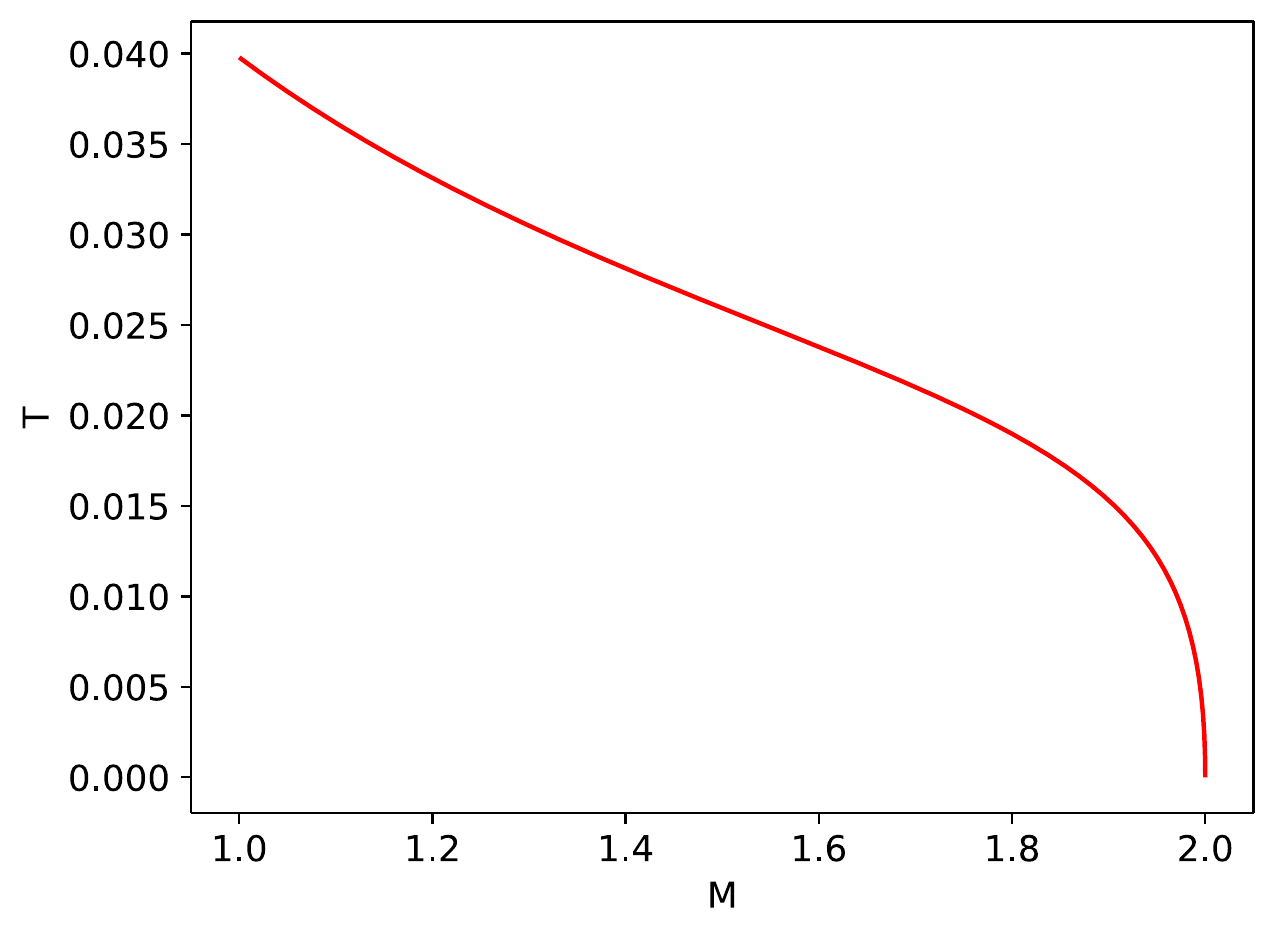}
        \caption{}
        \label{fig:Q_faster}
    \end{subfigure}
    \caption{%
        (a) Temperature evolution when mass increases faster than the charge-like parameter $N$ (here denoted $Q$). Starting from $Q = 0.01$, both $Q$ and $M$ grow, but $M$ increases at twice the relative rate. A clear phase transition appears at a critical mass, beyond which the temperature asymptotically approaches zero.
        (b) Temperature behaviour when $N$ (denoted $Q$) grows faster than mass. The temperature decreases monotonically with increasing $M$, driven predominantly by the rapid growth of $N$, which suppresses the Hawking temperature.
    }
    \label{fig:graphs M and Q}
\end{figure}

\begin{figure}[ht]
    \centering
    \includegraphics[width=0.6\linewidth]{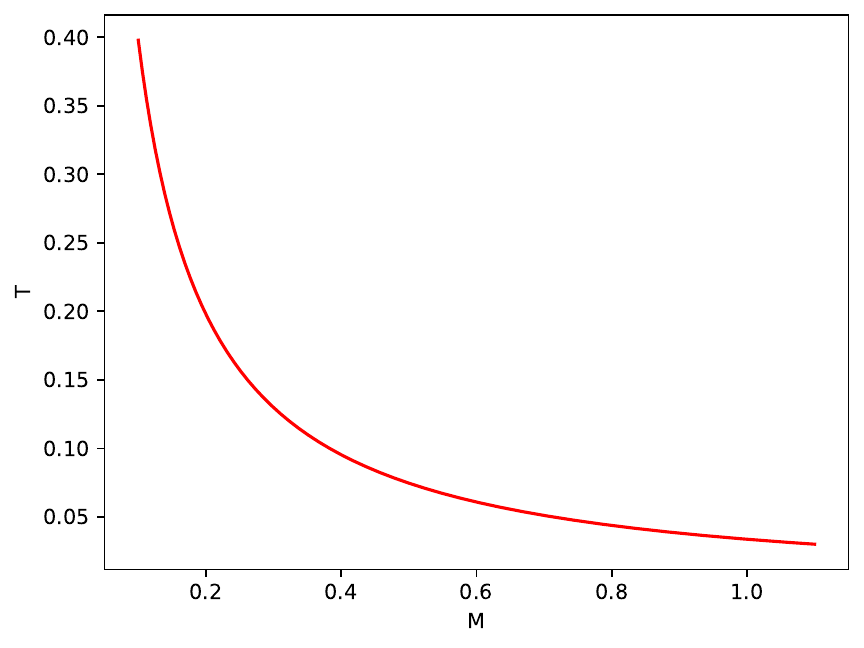}
    \caption{Temperature evolution under equal relative changes in $M$ and $N$ ($\beta = \gamma$). The curve exhibits intermediate behaviour between the two extreme cases, with no phase transition observed for the chosen parameters.}
    \label{fig:equal_changes}
\end{figure}
\section{Connection between Temperature Behaviour and Energy Conditions}
\label{sec:nec}
Previously, we considered the case of a static black hole. Although black hole parameters may change due to accretion or evaporation, we restricted our analysis to situations either before such processes began or after they had already concluded, thereby yielding a static spacetime metric. However, if we consider the dynamical process of accretion or evaporation, the appropriate metric takes the form
\begin{equation} \label{eq:tur}
ds^2 = -\left(1 - \frac{2M(v,r)}{r}\right) dv^2 + 2\,dv\,dr + r^2 d\Omega^2,
\end{equation}
where the mass function now depends not only on the radial coordinate $r$ but also on the advanced (Eddington–Finkelstein) time $v$. The generalization to the dynamical case leads to the dynamical Kiselev metric~\cite{bib:tur1, bib:tur2, bib:tur3}:
\begin{equation}
ds^2 = -\left(1 - \frac{2M(v)}{r} + \frac{N(v)}{r^{3\omega+1}}\right) dv^2 + 2\,dv\,dr + r^2 d\Omega^2.
\end{equation}
Here, $M(v)$ is the dynamical black hole mass, and for $\omega > 0$, the parameter $N(v)$ represents a combination of magnetic and electric charges sourced by nonlinear electrodynamics~\cite{bib:vertogradov2025non}. In the limit $N(v) \equiv 0$, we recover the Vaidya metric.

The metric~\eqref{eq:tur} is supported by matter whose energy-momentum content can be expressed in terms of the mass function $M(v,r)$ via Einstein's equations as follows:
\begin{eqnarray}
\mu &=& \frac{2\dot{M}}{r^2}, \nonumber \\
\rho &=& \frac{2M'}{r^2}, \nonumber \\
P &=& -\frac{M''}{r},
\end{eqnarray}
where $\mu$ denotes the energy flux density. The null energy condition (NEC) requires this flux to be non-negative at all times. However, during evaporation, this condition is violated. As shown in~\cite{bib:vertogradov2025plb}, such a violation leads to the following behavior: if apparent horizons lie within the region where energy conditions are violated, then the outer horizon is timelike and contracts, while the inner horizon is spacelike and expands. Conversely, when energy conditions are satisfied, the behavior is reversed-the outer horizon becomes a spacelike hypersurface and expands, whereas the inner horizon is timelike and contracts.

Additionally, a so-called \textit{energy-condition horizon} $r_{\text{nec}}$ was introduced, separating the spacetime region where energy conditions are violated from the region where they hold. This horizon is defined by the condition
\begin{equation}
\dot{M}(v, r_{\text{nec}}) = 0.
\end{equation}
It was also proven that an apparent horizon is null (i.e., a null hypersurface) if and only if it coincides with the energy-condition horizon $r_{\text{nec}}$. In this special case, the horizon remains stationary in time. We note that the fulfillment of the null energy condition is equivalent to $\dot{M} \geq 0$.

Given the dynamical nature of the Kiselev metric~\eqref{eq:tur}, we can explicitly determine the energy-condition horizon as
\begin{equation}
\dot{M} = 0 \quad \Rightarrow \quad r_{\text{nec}} = \left( \frac{\dot{N}}{2\dot{M}} \right)^{\frac{1}{3\omega}}.
\end{equation}
In the case of the charged Vaidya spacetime, where $\omega = \frac{1}{3}$ and $N = Q^2$, this yields the well-known result $r_{\text{nec}} = \frac{Q \dot{Q}}{2\dot{M}}$.

Thus, during mass evaporation with constant $N$, energy conditions are violated throughout the entire spacetime-as is the case for the standard Vaidya metric. Conversely, if the mass remains constant while the charge $N$ evaporates, energy conditions are satisfied everywhere, and we observe a contracting inner horizon and an expanding outer horizon.

However, the most interesting scenario occurs when both mass $M$ and charge $N$ evaporate simultaneously, leading to various possible horizon behaviors:
\begin{itemize}
\item The energy-condition horizon lies inside the inner horizon ($r_{\text{nec}} < r_-$). In this case, the null energy condition holds only in the region $0 \leq r \leq r_{\text{nec}}$. Consequently, the outer horizon contracts (timelike), while the inner horizon expands (spacelike). The black hole temperature decreases toward zero, as described in previous sections. This situation arises when mass and charge evaporate at approximately comparable rates.

    \item The energy-condition horizon is located between the two horizons ($r_- < r_{\text{nec}} < r_+$). Here, the NEC is satisfied for $0 \leq r \leq r_{\text{nec}}$, implying that both the outer and inner horizons contract. In this case, the temperature behaves analogously to the standard picture-diverging as the black hole mass becomes very small.

    \item The third case corresponds to charge evaporating much faster than mass. Here, the energy-condition horizon lies outside the outer horizon. Despite the decreasing mass, the outer horizon is an increasing function of time and asymptotically approaches $r_{\text{nec}}$, while the inner horizon continues to contract toward zero. At a certain moment, the outer horizon and the energy-condition horizon merge; thereafter, the outer horizon becomes a spacelike hypersurface and begins to shrink. In this scenario, the temperature initially decreases slightly and then diverges unboundedly, as in the standard black hole evaporation picture.
\end{itemize}
The case in which charge evaporates faster than mass is the most physically plausible, given that the dimensionless ratio of the electron charge to its mass is $\sim 10^{21}$. In this situation, the black hole should emit a flux of charged particles whose charge has the same sign as that of the black hole itself.
\section{Conclusion}
\label{sec:discussion} 
When considering the evaporation of a black hole possessing two horizons, it is essential not only to analyze thermodynamic quantities but also to examine the spacetime regions where the null energy condition (NEC) is violated. These regions play a pivotal role in the dynamical behavior of apparent horizons. In the standard evaporation picture-where only the black hole mass $M$ decreases-the NEC is violated throughout the entire spacetime. Consequently, the outer horizon contracts as a timelike hypersurface, while the inner horizon expands as a spacelike hypersurface.

As demonstrated explicitly for the Kiselev black hole, the Hawking temperature is intimately linked to the black hole's heat capacity: the temperature increases when the heat capacity is negative and decreases when it is positive. Near the extremal configuration, the heat capacity of the Kiselev black hole is always positive, ensuring that the temperature monotonically decreases to zero precisely at the moment the two horizons merge.

Thus, two distinct temperature behaviors emerge for the Kiselev black hole during evaporation:
\begin{enumerate}
    \item The temperature initially rises slightly (negative heat capacity), reaches a maximum at a thermodynamic phase transition, and then decreases to zero (positive heat capacity).
    \item If the heat capacity is already positive at the onset of evaporation, the temperature decreases monotonically to zero without exhibiting a maximum.
\end{enumerate}

The conventional Schwarzschild-like evaporation scenario-where temperature diverges as mass decreases-can only be recovered if both parameters $M$ and $N$ decrease simultaneously, with $N$ diminishing significantly faster than $M$. This process inevitably drives the black hole toward the Schwarzschild limit, restoring the standard evaporation picture. Such evolution occurs because the NEC is no longer violated everywhere in spacetime; in particular, the inner horizon may remain outside the NEC-violating region. 

Remarkably, this scenario can lead to counter intuitive behavior: during evaporation (i.e., decreasing $M$), the horizon radius may temporarily increase. However, this growth is not unbounded-once $N$ vanishes, the spacetime reduces to the Schwarzschild solution, and the horizon begins to shrink in the usual manner.

Since ref.~\cite{bib:vertogradov2025non} showed that for $\omega \in (0,1]$, the parameter $N$ corresponds to a combination of electric and magnetic charges supported by nonlinear electrodynamics, the reduction of $N$ must be accompanied by an outflow of charged particles from the black hole. Furthermore, according to the connection between black hole shadow dynamics and energy conditions established in~\cite{bib:ali2024plb}, the shadow radius in such an evaporation process would first increase and then decrease. This non-monotonic behavior could serve as an observational signature of black hole evaporation in future high-resolution imaging experiments.

\bibliography{ref}

\end{document}